\documentclass[showpacs,preprintnumbers,amsmath,amssymb, prd]{revtex4}
\usepackage{amsmath,amssymb,graphicx}
\usepackage {amssymb}
\newcommand{\nc}{\newcommand}
\nc{\ba}{\begin{eqnarray}}
\nc{\ea}{\end{eqnarray}}
\newcommand\be{\begin{equation}}
\newcommand\ee{\end{equation}}

\nc{\x}{{\bf{x}}}
\nc{\e}{{\bf{e}}}

\input{epsf.sty}

\begin{document}

\title{Issues on Generating Primordial Anisotropies  at the End of Inflation}

\author{Razieh Emami}
\email{emami@mail.ipm.ir}
\author{Hassan Firouzjahi}
\email{firouz@mail.ipm.ir}
\affiliation{School of Physics, Institute for Research in 
Fundamental Sciences (IPM),
P. O. Box 19395-5531,
Tehran, Iran}

\date{\today}

\begin{abstract}

We revisit the idea of generating primordial anisotropies at the end of inflation in models 
of inflation with gauge fields. To be specific we consider the charged hybrid inflation model where the waterfall field is charged under a $U(1)$ gauge field so the surface of end of inflation is controlled both by inflaton and the gauge fields.  Using $\delta N$ formalism properly we find that the anisotropies generated at the end of inflation from the gauge field fluctuations are exponentially suppressed on cosmological scales. This is because the gauge field evolves exponentially during inflation 
while in order to generate appreciable anisotropies at the end of inflation the spectator gauge field has to be frozen. We argue that this is a generic feature, that is, one can not generate observable anisotropies at the end of inflation within an FRW background. 

\end{abstract}

\vspace{0.3cm}

\preprint{IPM/P-2011/049}
\maketitle
\section{Introduction}

Primordial anisotropies captured considerable interests during past few years both observationally and theoretically. On the observational side there may be some indications of statistical anisotropies on cosmic microwave background (CMB)  \cite{Eriksen:2003db} although the statistical significances of these findings are under debate \cite{Komatsu:2010fb, Hanson:2009gu, Hanson:2010gu}. Motivated by these observations, there have been many attempts in the literature to generate primordial anisotropies during inflation. These models of inflation \cite{Ford:1989me, Kaloper:1991rw,  Kawai:1998bn,  Barrow:2005qv, Barrow:2009gx, Campanelli:2009tk, Golovnev:2008cf, Kanno:2008gn, Pitrou:2008gk, Moniz:2010cm, Boehmer:2007ut, Koivisto:2008xf, Golovnev:2011yc, Maleknejad:2011jr} usually require a gauge field or a vector field to seed the anisotropies at the order of few percent which may be detectable on CMB \cite{Ackerman:2007nb}, \cite{Yokoyama:2008xw}, \cite{Dimopoulos:2009vu, Dimastrogiovanni:2010sm, ValenzuelaToledo:2009af}. 

Models based on vector field suffer from ghost instability \cite{Himmetoglu:2008zp}
which makes the system unstable and physically unacceptable.  Therefore, it is crucial that the vector field is protected by a gauge symmetry so the longitudinal mode of the vector field excitations is not physical. On the other hand, because of the conformal invariance, the anisotropies generated during inflation from the quantum fluctuations of gauge field do not survive and are quickly damped on large scales by the end of inflation. Therefore it is essential that one breaks the conformal invariance while keeping the gauge symmetry explicit. This approach was employed in different contexts in  \cite{Martin:2007ue, Demozzi:2009fu,
Emami:2009vd,  Kanno:2009ei, Urban:2011bu}.
If one chooses the conformal factor in gauge kinetic term appropriately, the system can show an
attractor mechanism where the gauge field energy density becomes a sub-dominant but non-negligible component of total energy density \cite{Watanabe:2009ct, Emami:2010rm, Kanno:2010nr, Murata:2011wv, Bhowmick:2011em, Hervik:2011xm, Dimopoulos:2010xq}.
The amount of anisotropies generated are typically at the order of slow-roll parameters which can have important cosmological consequences. In these models where the generation of 
anisotropies is an attractor behavior of the system, the background explicitly breaks rotational 
invariance and instead of an FRW background one may have Bianchi I universe. As a result, the appearance of anisotropic fluctuations are a natural outcome of the system \cite{Gumrukcuoglu:2010yc, Dulaney:2010sq, Watanabe:2010fh, Pereira:2007yy}.
  
An interesting observation is made by Yokoyama and Soda \cite{Yokoyama:2008xw}
where primordial anisotropies may be generated at the end of inflation while the background
is still an FRW universe. The motivation is based on Lyth mechanism of generating curvature perturbations at the end of inflation \cite{Lyth:2005qk}. In Lyth formalism, the surface of end of inflation is controlled by a light scalar field, other than inflaton field, which can produce inhomogeneities at the end of inflation. For this mechanism to work, the additional scalar field has to be very light and scale invariant so it is frozen throughout inflation. 
With this motivation, Yokoyama and Soda considered a model where the surface of end of inflation is controlled by the  $U(1)$ gauge field $A_\mu$. As an specific model, the end of inflation can be as in hybrid inflation where now the waterfall field is gauged under $A_\mu$. It is argued in  \cite{Yokoyama:2008xw} that the quantum fluctuations of $A_\mu$ at the end of inflation can generate statistical anisotropies which can be observable. However, we shall show in this work that the requirement that  the additional spectator field  to be frozen throughout inflation do not apply for the gauge field. As a result, the anisotropies generated at the end of inflation are hugely suppressed due to conformal invariance. In other words, we show that one can not generate statistical anisotropies at the end of inflation 
from an FRW background. In order to generate appreciable anisotropies, one has to start with a background, like Bianchi I, where the rotational invariance is explicitly
broken.

The rest of paper is organized as follows. In section \ref{hybrid} we review the charged hybrid 
inflation model. In section \ref{power-spec} we perform the $\delta N$ formalism to calculate the curvature perturbations generated from the surface of end of inflation. 
The summary and discussions are in section
\ref{summary}.   We relegate the technical details  of $\delta N$ formalism into Appendix \ref{app-anisotropy}.


\section{Charged Hybrid Inflation}
\label{hybrid}

In this section we study charged hybrid inflation model in some details
which serves as a set up to study anisotropies generated at the end of inflation as in Yokoyama and Soda studies. The model is based on the action \cite{Emami:2010rm}
\ba
\label{action3} S=\int d^4 x  \sqrt{-g} \left [ \frac{M_P^2}{2} R - \frac{1}{2} \partial_\mu \phi
\,  ^\mu\phi- \frac{1}{2} D_\mu \psi
\,  D^\mu\bar\psi - \frac{f^{2}(\phi)}{4} F_{\mu \nu} F^{\mu
\nu}- V(\phi, \psi, \bar \psi) \right]  \, .
\ea
where $\phi$ is the inflaton field while $\psi$ is the complex waterfall field. The covariant derivative is defined via
\ba
D_\mu \psi = \partial_\mu  \psi + i \e \,  \psi  \, A_\mu
\ea
where $\e$ is
the dimensionless gauge coupling of $A_\mu$ to $\psi$. As usual, the
gauge field strength is given by
\ba F_{\mu \nu} = \nabla_\mu A_\nu
- \nabla_\nu A_\mu  = \partial_\mu A_\nu - \partial_\nu A_\mu \, .
\ea
As mentioned in the introduction the conformal factor $f(\phi)$ with an appropriate form
is added in order to break the conformal invariance so the gauge field energy density and its quantum fluctuations do not dilute during inflation. 

We are interested in configurations where the potential is axially symmetric and $V(\psi, \bar \psi , \phi)= V(\chi, \phi)$ where $ \psi(x) = \chi(x) \,  e^{i \theta(x)}$. The potential is as in standard hybrid inflation \cite{Linde:1993cn}
\ba
\label{pot3} V(\phi, \chi)=\frac{\lambda}{4 }  \left(  \chi^2 - \frac{M^2}{\lambda} \right)^2 + \frac{g^2}{2} \phi^2 \chi^2 + \frac{m^2}{2} \phi^2 \, .
\ea

In general in the presence of background gauge field, the system will lose the rotation invariance. In particular, we can take the background gauge field to have the form $A_{\mu}=(0,A(t),0,0)$
and the background will be Bianchi type I universe. If the conformal coupling $f(\phi)$
is chosen appropriately, the system reaches an attractor mechanism in which the gauge field energy density becomes sub-leading but nonetheless non-negligible  compared to 
total energy  \cite{Watanabe:2009ct}.
As demonstrated in  \cite{Emami:2010rm} the background anisotropies are at the level of slow-roll parameters. In principle one has to consider the cosmological perturbation analysis where the background is not an FRW universe. The cosmological perturbations analysis were performed explicitly in \cite{Gumrukcuoglu:2010yc, Dulaney:2010sq, Watanabe:2010fh}
for the chaotic model studied in  \cite{Watanabe:2009ct}. We shall come back to detailed cosmological perturbations analysis in the charged hybrid inflation model with Bianchi I background elsewhere. Instead, here we follow the logic in \cite{Yokoyama:2008xw} where the background gauge field
does not destroy the rotational invariance so inflation proceeds as in standard FRW background. Our aim is to see whether or not one can generate statistical anisotropies at the end of inflation from the gauge field quantum fluctuations.

The details of background fields equation in Bianchi I background were studied in \cite{Emami:2010rm}. Here we borrow only the important results.  At the background level, we start with the FRW metric which has the following form 
\ba
\label{metric} ds^2 = - dt^2 + a(t)^2 d \x ^2 \, .
\ea
The total energy density in the slow-roll limit where the kinetic energies of $\phi$ and $\chi$
are negligible is obtained to be

\ba
\label{energy}
{\cal E}=V(\phi,\chi) + e^{-2N(t)} \left( \frac{1}{2}f^2(\phi) \dot
A^2+\frac{\mathbf{e}^2\chi^2}{2} A^2 \right) \, ,
\ea
where $N(t)$ is the number of e-foldings defined as $ d N=H dt $ with $H =\dot a/a$ to be the Hubble expansion rate during inflation.  As can be seen from Eq. (\ref{energy}) the gauge field has two contributions in total energy density, from its kinetic energy and from its contribution to potential energy with the coupling $\e^2 \chi^2$. It is this latter contribution which is essential for our studies below.

The interesting new effect is that the gauge coupling $\mathbf{e}$ induces a new time-dependent mass term for the waterfall field in the form $\mathbf{e}^2 e^{-2 N} A^2 \,  \chi^2$.  As in standard hybrid inflation \cite{Linde:1993cn, Copeland:1994vg}
we work in the vacuum dominated regime where the waterfall field is very heavy during inflation so $\chi$ quickly settles down to its instantaneous minimum $\chi=0$ during inflation. In standard hybrid inflation  models, inflation ends when  inflaton field reaches a critical value,
$\phi=\phi_c \equiv \frac{M}{g}$, where the waterfall field becomes tachyonic and rolls down very quickly to its global minimum $\psi=\mu\equiv M/\sqrt{\lambda}, \phi=0$ ending inflation very efficiently. In the current model due to coupling of gauge field to waterfall field, the surface of end of inflation is modified. Calculating the effective mass of waterfall we have
\ba
\label{chi-mass}
\frac{\partial^2 V}{\partial \chi^2}\large|_{\chi=0}  = g^2 (\phi^2 - \phi_c^2) + \mathbf{e}^2 e^{- 2 N} A^2 \, .
\ea
 In the absence of the gauge field, the onset of waterfall instability is when $\phi= \phi_c$. However, in the presence of gauge field the time of waterfall transition is modified.
 
The condition of waterfall phase transition, Eq. (\ref{chi-mass}), can be rewritten as
\ba
\label{transition}
\hat \phi^2 + \frac{\mathbf{e}^2}{g^2} \hat A^2  =1
\ea
where we have defined the dimensionless fields 
\ba
\label{hat-fileds}
\hat \phi \equiv \frac{\phi}{\phi_c} \quad , \quad \hat A \equiv \frac{A}{\phi_c} \, .
\ea
In this notation, in the absence of gauge field the waterfall phase transition happens at $\hat \phi=1$.  

Note that we have chosen the convention that the time of end of inflation corresponds to $N\equiv N_f=0$ and count the number of e-foldings backward. In order to solve the flatness and the horizon problem we require  inflation lasts for at least 60 e-foldings so at the start of inflation $N\equiv N_i \simeq-60$.

As mentioned above, we assume the waterfall field is very heavy during inflation and the potential driving inflation is
\ba
V\simeq \frac{M^4}{4\lambda}+\frac{1}{2}m^2\phi^2  \, .
\ea
In order for the inflaton field to be light during inflation so the slow-roll conditions are met we need $p_c\gg1$ where $p_c$ is defined via
\ba
p_c \equiv \frac{M^4}{2 \lambda m^2 M_P^2} \, .
\ea
Furthermore, the assumption that the waterfall field is very heavy during inflation requires
$\lambda M_P^2 /M^2 \gg1$. Also, the condition of vacuum domination during inflation
is met if $\lambda/g^2 \ll M^2/m^2$. Finally, we work in the limit where the waterfall phase transition is very sharp so inflation ends abruptly in less than an e-fold. For this to happen
one requires that $\lambda M_P^2 /M^2 \gg p_c$ \cite{Abolhasani:2010kr}.

Considering the vacuum dominated potential and assuming the gauge field has no appreciable contribution  to energy density the inflaton dynamics is
\ba
\label{phase1-hybrid}
\phi' + \frac{2 \phi}{p_c} =0 \rightarrow  \phi \simeq \phi_{f} e^{-2 N /p_c}  \, ,
\ea
where $\phi_{f}$ is the final value of the inflaton field and here and below, a prime indicates the derivative with respect to number of e-fold. Alternatively,  the number of e-foldings can be written as
\ba
\label{N-phi-hybrid}
N(\phi) \simeq -\frac{p_c}{2} \ln \left( \frac{\phi}{\phi_{f}} \right)  \, .
\ea
As mentioned before, we count $N$ backwards so we normalize $N$ such that 
at the end of inflation $N=N_f=0$ while at the start of inflation where $\phi=\phi_i$, we have $N=N_i \simeq-60$.

In order to ensure the background is and FRW universe, the gauge field energy density should be very small. During inflation when $\chi=0$ the third term in Eq. (\ref{energy}) basically vanishes. Therefore, in order to keep the background isotropic, 
we have to make sure the gauge field kinetic energy is always sub-leading during inflation. This is basically controlled by the form of conformal factor $f(\phi)$.
It is useful to define the dimensionless coupling $R$ which measures the fraction of energy stored in gauge field kinetic energy
\ba
\label{R}
R \equiv \frac{\dot A^2 f(\phi)^2 e^{-2N}}{2 V} \, .
\ea
One can easily check from the gauge field equation that \cite{Emami:2010rm}
\ba
\label{A-prime}
\left (f(\phi)^2 e^N A' \right)' =0
\ea
so $A' \sim e^{-N} f(\phi)^{-2}$.  Consequently, $R$ scales like $R \sim f(\phi)^{-2} e^{-4 N} \sim
f(\phi)^{-2} \phi^{2 p_c}$. Therefore with the conformal factor in the form 
$f (\phi) = \hat \phi^p$ with $p>p_c$, the gauge field energy density reaches the attractor mechanism and sometime during inflation $R$ becomes comparable to the slow-roll parameters.
However, for $p<p_c$ the gauge field energy density is diluted due to conformal invariance.
The special case $p=p_c$ is interesting in which $R$ remains fixed and as we shall see below the modified gauge field $\delta { B} \equiv f \delta A$ becomes scale invariant.  Therefore,  we consider the case where $p=p_c$ and 
\ba
\label{f2}
f(\phi) \propto   \phi^{p_c} \propto e^{-2 N}  \propto a^{-2} \, .
\ea
To fix the overall numerical normalization of $f(\phi)$ we note that $f(\phi)^{-1}$ measures the effective perturbative gauge kinetic coupling so in order to keep the gauge theory under perturbative control we require that $f(\phi)$ is large. We assume that during inflation $f(\phi)$
is exponentially large so the gauge theory is perturbatively under control. As inflation proceeds
$f(\phi)$ decreases exponentially as can be seen from Eq. (\ref{f2}). It should be such that towards the end of inflation, $f(\phi)$ reaches its final value $f(\phi_f) \gtrsim 1$ so the the gauge theory is still perturbatively under control.  For future reference, we note that 
\ba
\label{f3}
f(\phi) = f(\phi_f) e^{-2 N} \, .
\ea

With gauge kinetic coupling given by Eq. (\ref{f2}), the gauge field equation (\ref{A-prime})
can be solved easily and
\ba
\label{A-prime-hybrid}
\hat A = \hat A_f + \kappa \left( e^{3N}-1 \right) \, ,
\ea
where $\kappa$ is a constant of integration and $\hat A_f$ is the value of the gauge field at the end of inflation. Alternatively, one can find the number of e-foldings in terms of the gauge field via
\ba
\label{N-A-hybrid}
N= \frac{1}{3} \ln \left( \frac{\hat A - \hat A_{f} + \kappa}{\kappa} \right) \, .
\ea

From Eq. (\ref{A-prime-hybrid}) we find that the gauge field is essentially negligible at the start of inflation, $N_i\simeq-60$, and grows exponentially towards the end of inflation where it approaches $\hat A_f \simeq \kappa$. The exponential growth of the gauge field towards the end of inflation is the key factor in our discussion below in determining the anisotropies induced at the end of inflation. However, there is a bound on how large $A$ can be at the end of inflation. From Eq. (\ref{transition}) we see that $(\e/g)\hat A_f <1$ in order to
terminate inflation. Taking $\hat A_f \simeq \kappa$, this in turn yields $\kappa < g/\e$.

In order to make sure that the exponential growth of the gauge field does not destroy the isotropic FRW background, we have to impose the condition that the gauge field fraction of energy, $R$ defined in Eq. (\ref{R}),  is smaller than the slow roll parameter $\epsilon = \frac{M_P^2}{2} (\frac{V_{,\phi}}{V})^2$.
As shown in \cite{Emami:2010rm}, if the gauge field energy density increases such that $R\sim \epsilon$, then the system reaches the attractor regime where the background is a Bianchi I universe and one can not neglect the anisotropies in fields equations, mainly the back-reaction of gauge field on inflaton dynamics. In order to prevent this from happening so our background is an FRW universe all the way till end of inflation, we have to impose the condition $R< \epsilon$.
This in turn yields
\ba
\label{kappa}
\kappa f(\phi_f)  < \sqrt{\frac{4}{3}}
 \frac{ \phi_i}{p_c \phi_c}
\ea
As an order of magnitude estimate, with $\phi_i \sim \phi_c$ and $f(\phi_f) \sim 1$, one conclude that $\kappa < 1/p_c$.

To calculate the power spectrum and bispectrum we use the powerful $\delta N$ formalism 
\cite{Sasaki:1995aw, Wands:2000dp, Lyth:2005fi}. Our goal is to  calculate $\delta N(\hat \phi,\hat A)$ analytically to second order in terms of  $\delta \hat \phi$ and $\delta \hat A$.
In \cite{Yokoyama:2008xw} they employed the mechanism  in \cite{Lyth:2005qk} where it was assumed that there are two different contributions in $\delta N$. The first one comes from the evolution of the {\it inflaton} field during the inflation and the second one originates from the fluctuations of {\it gauge field } at the end of inflation. In order to use  Lyth mechanism directly, the additional field other than inflaton (in our case the gauge field $A_\mu$) should be very light and scale invariant. As mentioned before, it is the re-scaled gauge field perturbations $\delta { B} = f \delta A$ which is scale invariant. On the other hand, it is the gauge field $A$ and not the re-scaled field $B$ which appears in Eq. (\ref{transition})  governing the surface of end of inflation. As a result, it is not clear if one can use the mechanism in \cite{Lyth:2005qk} automatically for the case at hand. In order to prevent any confusion, we calculate $\delta N$ directly from first principle. Our method is similar to multi-brid analysis employed by Sasaki and Sasaki-Naruko in \cite{Sasaki:2008uc, Naruko:2008sq}, see also \cite{Huang:2009vk}.

Now we return to the surface of the end of inflation given by Eq. (\ref{transition}). As in the  \cite{Sasaki:2008uc, Naruko:2008sq}, it is convenient to introduce the angle $\gamma$ such that
 \ba \label{gamma}
\hat \phi_{f} = \cos \gamma \quad , \quad
\hat A_{f} = \frac{g}{\mathbf{e}} \sin \gamma \, . 
\ea
Our goal is to calculate $\delta N$ as a function of $\delta \phi$ and $\delta A$ up to second order, taking into account the fluctuation $\delta \gamma$ generated at the surface of end of inflation.  We relegate the details of the analysis to
 Appendix \ref{app-anisotropy}. Calculating $\delta N(\hat \phi,\hat A)$ up to second order
we have

\ba
\label{deltaN}
-\frac{2}{p_{c}}\delta N(\hat \phi,\hat A) = { \cal A}~\frac{\delta \hat \phi}{\hat \phi} + {\cal N}~\delta \hat A + {\cal I}~ \left(\frac{\delta \hat \phi}{\hat \phi}\right)^{2} +  {\cal S}~\left(\delta \hat A\right)^{2} + {\cal T}~\left( \frac{\delta \hat \phi}{\hat \phi}\delta \hat A \right)
\ea
where 
\ba
{\cal A}&&\equiv \left( \frac{{g \cos{\gamma}}}{3\kappa \mathbf{e}} \frac{ e^{-3N}}{Y}\right) \nonumber\\
{\cal N}&&\equiv \left( \frac{{\tan{\gamma}}}{3\kappa} \frac{ e^{-3N}}{Y}\right) \nonumber\\
{\cal I}&&\equiv \left( \frac{p_{c}^{2} g}{24 \mathbf{e}\kappa} (\frac{1 + 4 \sin^2{\gamma}}{\cos{\gamma}}) \frac{ e^{-3N}}{Y^{3}}+ \frac{p_{c} g^2 }{12 \mathbf{e}^{2}\kappa ^{2}}\sin {2\gamma} (\frac{1}{3} - \frac{p_{c}}{4})\frac{ e^{-6N}}{Y^{3}} - \frac{g^3}{54 \mathbf{e}^{3}\kappa ^{3}} \cos^{3}{\gamma} \frac{ e^{-9N}}{Y^{3}}\right) \nonumber\\
{\cal S}&&\equiv \left( -\frac{p_{c}^{2}}{24 \kappa^2}\tan^3\gamma \frac{ e^{-6N}}{Y^{3}}+ \frac{g}{54 \mathbf{e} \kappa ^{3}} (\frac{1 +  \sin^2{\gamma}}{\cos{\gamma}}) \frac{ e^{-9N}}{Y^{3}}\right) \nonumber\\
{\cal T}&&\equiv \left( \frac{p_{c} g}{6\mathbf{e} \kappa^2}\left[\frac{((\frac{1}{3} - \frac{p_{c}}{2})\sin^2{\gamma} + \frac{1}{3})}{\cos{\gamma}}\right] \frac{e^{-6N}}{Y^{3}}\right) \nonumber\\
\,
\ea
The parameter $Y$ that appeared frequently in these formula  is given by
\ba
Y\equiv  \frac{g \cos{\gamma}}{3\kappa \mathbf{e}}e^{-3N} - \frac{p_{c}\tan{\gamma}}{2}
\ea


\section{Power Spectrum and Bispectrum}
\label{power-spec}

Now we compute the curvature perturbation ${\cal P}_{\zeta}(k)$ and the magnitude of non-Gaussianity parameter $f_{NL}$ in our model.

\subsection{Power Spectrum}

As usual we assume that the scalar field fluctuations  in Fourier space, $\delta  \phi_{k}$, are Gaussian with the dispersion relation 
\ba
\label{dis phi}
\left<\delta  \phi_{k} \delta \phi_{k'}\right> \equiv (2\pi)^{3} P_{\delta  \phi}(k)~\delta^3(k+k') ~~,~~ {\cal P}_{\delta  \phi}\equiv \frac{k^{3}}{2 \pi^{2}}P_{\delta  \phi}(k)\,
\ea
So the power spectrum of $\hat \phi = \phi/\phi_c$ is 
\ba
\label{power phi}
{\cal P}_{\delta \hat \phi} = \left. \left(\frac{H}{2\pi \phi_c}\right)^{2}\right|_{*} 
\delta^3(k+k') \,
\ea
It is important to not that this is calculated at the time of horizon crossing, denoting by an $*$, 
when $k=a_* H$.

To find the power spectrum of the gauge field $A_\mu$, we have to first look into its quantum fluctuations. Choosing the Coulomb-radiation gauge where $\delta A_0 = \nabla. A=0$ and defining the rescaled perturbations 
\ba
\label{rescale-A}
\delta {B}_k \equiv f(\phi) \delta A_k
\ea
the fluctuations of the gauge field is given by
\ba
\label{schrodinger}
{\delta {B}_k}'' + \left( k^2 - \frac{f''}{f}
\right) \delta {B}_k =0
\ea
where the derivative here is with respect to the conformal time $d \tau = -dt/a(t)$. As advertised before, if we take $p=p_c$ so $f(\phi) \propto \hat \phi^{p_c} \propto a(\tau)^{-2}$, then 
$\frac{f''}{f} = \frac{2}{\tau^2}$ and the power spectrum of $\delta {B}_k$ is scale invariant.
If we take $p>p_c$, but still such that $R < \epsilon$ and 
FRW is a good approximation to the background, 
then $\delta B$ excitations will become mildly scale dependent  but it will not affect our main conclusion below. As a result, with $p=p_c$, we have \cite{Yokoyama:2008xw}
\ba
\label{power A} 
{\cal P}_{\delta \hat A} = f(\phi)^{-2} {\cal P}_{\delta \hat \phi}   \, .
\ea
Note that this relation always hold during inflation. In particular, at the time of horizon crossing
for the mode of cosmological interest  $k=a_* H$ we have 
${\cal P}_{\delta \hat A} = f(\phi_*)^{-2} {\cal P}_{\delta \hat \phi}  \sim e^{4N_i} {\cal P}_{\delta \hat \phi}$. For $N_i \simeq-60$, one concludes that ${\cal P}_{\delta \hat A}$ is completely suppressed compared to ${\cal P}_{\delta \hat \phi}$. As we shall see below, this is the key effect in suppressing the anisotropies generated from the surface of end of inflation. 

Having obtained the power spectrum of $\delta \hat \phi_k$ and $\delta \hat {B}_k$
we are able to calculate the power spectrum of the curvature perturbation $\zeta_k$. Using the 
$\delta N$ prescription 
\ba
\zeta(\mathbf{x},t) &&= \delta N(\hat \phi,\hat A) \ea
the power spectrum of $\zeta_k$ is obtained to be
\ba
\label{power}
{\cal P}_{\zeta} &=& \left(\frac{p_{c} \cal A}{2 \hat \phi}  \right)_*^2 
{\cal P}_{\delta \hat \phi}  \left[ 1+ f(\phi_*)^{-2}
\left(\frac{\cal N \hat \phi}{\cal A}\right)_*^2
\right]  \nonumber\\
&\equiv&{\cal P}^{(0)}_{\zeta} \left[ 1+ \frac{\Delta {\cal P}_{\zeta}}{{\cal P}^{(0)}_{\zeta}}
\right]
\ea
where ${\cal P}^{(0)}_{\zeta}=  \left(\frac{p_{c}  \cal A}{2 \hat \phi}  \right)_*^2 
\times \left(\frac{H}{2 \pi \phi_c}\right)^2
= \left(\frac{H  p_c}{2 \phi_*}  \right)^2
$
represents the curvature perturbations originated from the inflaton field at the time of horizon crossing in the absence of gauge field when $\e=0$. 
It is crucial to note that all dynamical quantities above are calculated at the time of horizon crossing when $N_*\simeq -60$. Now we can look at the anisotropies generated at the end of inflation which is encoded in the the second term in big bracket in Eq. (\ref{power}). Specifically, we have
\ba
\label{delta-power}
\frac{\Delta {\cal P}_{\zeta}}{{\cal P}^{(0)}_{\zeta}} = \frac{1}{f(\phi_*)^2} \left( \frac{\e\,  \hat \phi_* \sin \gamma}{g\,   ({\cos \gamma})^2}
\right)^2 =  e^{4N_*}
\left( \frac{\e\,  \hat \phi_* \sin \gamma}{g\,  f(\phi_f)  ({\cos \gamma})^2}
\right)^2 \, ,
\ea
where in the last equation Eq. (\ref{f3}) has ben used. As mentioned earlier, to keep the gauge theory under perturbative control we require 
$ f(\phi_f) \gtrsim 1$. For any reasonable  values of $\e/g, \gamma$, not exponentially different from unity,  and with 
$\hat \phi_* = \phi_*/\phi_c \sim1$, we conclude that the induced anisotropy scales like $ e^{4N_*} \sim e^{-240}$ and is completely suppressed on cosmological scales. The situation is somewhat similar to the effects of waterfall quantum fluctuations in standard hybrid inflation as studied recently in \cite{Abolhasani:2010kr, Hybrid} where  it is found that the induced curvature perturbations from  waterfall quantum fluctuations 
scales like $e^{3 N_*}$ and is  completely suppressed on cosmological scales. 

Having this said, one may wonder why the results obtained here is drastically different from the results obtained in \cite{Yokoyama:2008xw}. As we commented earlier, the analysis in 
\cite{Yokoyama:2008xw} relies on Lyth mechanism of generating curvature perturbation at the end of inflation \cite{Lyth:2005qk}. However, as we have seen, the basic assumptions to
employ the mechanism in \cite{Lyth:2005qk} are violated here. In order to borrow the mechanism in \cite{Lyth:2005qk} directly, the spectator gauge field should frozen during inflation. The surface of end of inflation, given by 
Eq.(\ref{transition}),
is controlled by the standard field $A_\mu$.  However, as we see from Eqs. (\ref{rescale-A}) and (\ref{schrodinger}), it is the re-scaled field $\delta {B} = f \delta A$ which plays the role of the light and scale invariant field so the roles of $\delta A$ and $\delta {B}$ are mixed. Alternatively, we see from Eq. (\ref{A-prime-hybrid}) that
the gauge field $A$ is evolving exponentially and it obviously violates the requirement that 
$A$ to be light, i.e. frozen, as required  in \cite{Lyth:2005qk}.

\subsection{Bispectrum}
Here we would like to calculate the non-gaussianity parameter $f_{NL}$in this model.
As usual, the bispectrum is defined via
\ba
\label{bispectrum}
\left< \zeta_{k} \zeta_{k'} \zeta_{k''}\right> \equiv (2\pi)^{3}B_{\zeta}(k,k',k'') \delta^{3}(k+k'+k'')
\ea
Correspondingly, $f_{NL}$ is given by
\ba
\frac{6}{5} f_{NL} \equiv \frac{B_\zeta(k, k', k'')}{P_\zeta(k)  P_\zeta(k') + c.p.} \, .
\ea
Since in our model the contribution of the scalar field in the power spectrum, 
${\cal P}^{(0)}_{\zeta}$, is dominant we can approximate the denominator above by
the isotropic part of power spectrum. Our goal is to compare different contributions to 
$f_{NL}$ mainly thorough their $N_*$ dependence. 

At the tree level, there are three sources of non-Guassianity form the interactions 
${\cal A}^2 {\cal I} \delta \phi^4   $, ${\cal N}^2 {\cal S} \delta A^4$ and 
${\cal A} {\cal N} {\cal T} \delta \phi^2 \delta A^2$. Furthermore, there are four more sources 
of non-Gaussianities at the loop level which are suppressed compared to tree level. Now compare different sources of non-Gaussianities at tree-level. Denoting the contributions of these interactions in $f_{NL}$ by $f_{NL}^{(\phi)}$,  $f_{NL}^{(A)}$ and  $f_{NL}^{(\phi-A)}$ respectively, we have
\ba
f_{NL}^{(\phi)} \simeq \frac{2}{p_c}
\quad , \quad
\ea
\ba
\frac{f_{NL}^{(A)}}{f_{NL}^{(\phi)}} \sim 
\frac{{\cal N}^2 {\cal S}}{{\cal A}^2 \cal I} f(\phi_*)^{-4} \simeq e^{8 N_*} \, ,
\ea
and
\ba
\frac{f_{NL}^{(\phi-A)}}{f_{NL}^{(\phi)}} \sim 
\frac{{\cal N} {\cal T}}{{\cal A} \cal I} f(\phi_*)^{-4} \simeq e^{7 N_*}\, .
\ea
As expected, the isotropic non-Gaussianity, $f_{NL}^{(\phi)}$, is at the order of slow-roll parameter as in single field models of inflation. Interestingly, with $N_* \simeq -60$,
we see that the anisotropic non-Gaussianities are completely suppressed compared to $f_{NL}^{(\phi)}$. 

\section{Summary and Discussions}
\label{summary}

In this work we have revisited the question of generating primordial anisotropies at the end of inflation.  To be specific we considered the charged hybrid inflation scenario where the waterfall field is charged under the $U(1)$ gauge field. Because of the interaction 
$\e A^2 \chi^2$ the onset of waterfall instability is controlled both by gauge field and the inflaton field.  We worked in the limit of a very sharp waterfall phase transition so inflation ends abruptly after the waterfall.

Calculating curvature perturbations carefully using $\delta N $ formalism, following the 
methods in \cite{Sasaki:2008uc, Naruko:2008sq}, we  have shown that the anisotropies generated at the end of inflation, both in power spectrum and bispectrum,  are exponentially suppressed on cosmological scales. This is in contrast with the results obtained in  \cite{Yokoyama:2008xw}. The analysis in \cite{Yokoyama:2008xw} relies on Lyth mechanism \cite{Lyth:2005qk} where  a spectator field, field other than the inflaton field, controls the surface of end of inflation. If this spectator field is very light, then it is frozen during inflation and its fluctuations at the end of inflation generate inhomogeneities in 
$\delta N$. However, as we have shown, in order to keep the dynamics of gauge field relevant to affect the surface of end of inflation, one has to add the conformal factor $f(\phi) \sim e^{-2N}$
such that the gauge field evolves exponentially during inflation. This clearly violates the criteria that the gauge field to be frozen as required in \cite{Lyth:2005qk}. Putting it another way, it is the fluctuation
$\delta {B} = f \delta A$ which is frozen and scale invariant after horizon exit, while the surface of end of inflation is controlled by $A$. This mixes the roles of $\delta A$ and $\delta {B}$ and one can not borrow the mechanism in \cite{Lyth:2005qk} directly for the case at hand.

Performing $\delta N$ analysis carefully, we have demonstrated that the induced anisotropies scale like $e^{4 N_*}$ where  $N_* \simeq -60$. We also showed that the anisotropies in 
$f_{NL}$ are even more suppressed. This clearly shows that there are no anisotropies generated at the surface of end of inflation in this model where the background is an FRW universe. 

Having this said, one may wonder how generic this conclusion is. We argue that this result is generic and does not rely on the particular model of charged hybrid inflation which we studied here.
The reason is that, in order to keep the gauge field fluctuations to survive during inflation and affect the surface of end of inflation, one has to break the conformal invariance by a time dependent gauge kinetic coupling. However, it is $\delta {B}$ which is frozen and scale invariant while the the background gauge field is exponentially increasing due to conformal factor. As a result  the spectrum of $\delta A$ is exponentially suppressed at the time when cosmological scales leave the horizon. However, this argument does not apply to models where the isotropy is explicitly broken at the background, such as in Bianchi I universe, studied
in \cite{Watanabe:2009ct, Emami:2010rm, Kanno:2010nr}. In theses models, since the gauge field fluctuations are not suppressed at the time of horizon crossing, one indeed expects to find non-negligible primordial anisotropies as explicitly studied in \cite{Gumrukcuoglu:2010yc, Dulaney:2010sq, Watanabe:2010fh}.

\section*{Acknowledgement}
We would like to thank Ali Akbar Abolhasani, Thorsten Battefeld, 
David Lyth, Mohammad Hossein Namjoo,
Misao Sasaki, Jiro Soda and Shuichiro Yokoyama for helpful discussions, critical comments and correspondences.  
\appendix{}


\section{Evaluation of $\delta N $     for quadratic potentials   }
\label{app-anisotropy}

Here we provide the calculations resulting in Eq. (\ref{deltaN}) in some details.
First of all, let us introduce the Taylor expansion of a two-variable functional , $ F = F(\Phi,\Psi(\gamma))$. Setting $\delta \gamma = \delta _{1} \gamma + \delta _{2} \gamma $, where $ \delta _{1} \gamma $ and  $ \delta _{2} \gamma $ are linear and second order perturbations respectively, the perturbation of the $F$ to second order is
\ba
\label{taylor sery}
\Delta F && = \frac {\partial F}{\partial \Phi}\delta \Phi + \frac {\partial F}{\partial \Psi} \left(\frac {\partial \Psi}{\partial \gamma} (\delta_{1} \gamma + \delta_{2} \gamma) + \frac{1}{2} \frac {\partial^{2} \Psi}{\partial \gamma ^{2}}(\delta_{1} \gamma)^{2} \right) + \frac{1}{2} \frac {\partial^{2} F}{\partial \Phi ^{2}} (\delta \Phi)^{2} + \frac{1}{2} \frac {\partial^{2} F}{\partial \Psi ^{2}} \left( \frac {\partial \Psi}{\partial \gamma}\right)^{2} (\delta_{1} \gamma)^{2} \nonumber\\
&&~~ + \frac {\partial^{2} F}{\partial \Phi \partial \Psi } \frac {\partial \Psi}{\partial \gamma} \left( \delta \Phi  \delta_{1} \gamma \right) \,
\ea
Now recall $N$ as a function of fields given  in Eq. (\ref{N-phi-hybrid}) and Eq. (\ref{N-A-hybrid}),
\ba
\label{N-phi-A}
N &&= -\frac{p_c}{2} \ln \left( \frac{\hat \phi}{\hat \phi_{f}} \right) \nonumber\\
&&= \frac{1}{3} \ln \left( \frac{\hat A - \hat A_{f} + \kappa}{\kappa} \right) \, ,
\ea
in which $\hat \phi_{f}$ and $\hat A_{f}$ are parameterized in terms of $\gamma$ as in Eq. (\ref{gamma}),
\ba
\label{re-gamma}
\hat \phi_{f} = \cos \gamma \quad , \quad
\hat A_{f} = \frac{g}{\mathbf{e}} \sin \gamma \, .
\ea
Then, one can calculate $\delta N$ either from the left or right hand side of the Eq. (\ref{N-phi-A}) to second order
\ba
\label{delta N phi}
\delta N &&= - \frac{p_{c}}{2}\left( \frac{\delta \hat \phi}{\hat \phi} - \frac{1}{2}\left(\frac{\delta \hat \phi}{\hat \phi} \right)^{2} + \tan \gamma (\delta_{1}\gamma+\delta_{2}\gamma) + \frac{1}{2\cos^{2}\gamma}(\delta_{1}\gamma)^{2}\right)\nonumber\\
~~~&&= \frac{e^{-3 N}}{3\kappa}\delta \hat A - \frac{e^{-6 N}}{6\kappa^2}(\delta \hat A)^{2}- \frac{g \cos
\gamma}{3\mathbf{e}\kappa}e^{-3 N}\left(\delta_{1}\gamma + \delta_{2}\gamma\right) \nonumber\\
 ~~~&&+ \left(\frac{g\sin \gamma}{6\mathbf{e}\kappa}e^{-3 N}-\frac{g^{2}\cos^{2}\gamma}{6\mathbf{e}^{2}\kappa^{2}}e^{-6 N}\right)(\delta_{1}\gamma )^{2} + \frac{g \cos \gamma}{3\mathbf{e}\kappa^{2}}e^{-6 N} (\delta_{1}\gamma \delta \hat A) \nonumber\\
\ea
The linear part of the above equation determines $\delta_{1} \gamma$. We find
\ba
\label{delta1}
~~~~~~~~~~~~~~\delta_{1} \gamma = \left. \left(\frac{p_{c}}{2}\frac{\delta \hat \phi}{\hat \phi} + \frac{e^{-3 N}}{3\kappa}\delta \hat A\right) \right / \left( \frac{g \cos \gamma}{3e\kappa}e^{-3 N} - \frac{p_{c}}{2}\tan \gamma \right)
\ea
Then collecting the second order terms in Eq. (\ref{delta N phi}) ,we find
\ba
\label{delta2}
\delta_{2} \gamma = \left(\mathbf{a} \left(\frac{\delta \hat \phi}{\hat \phi}\right)^{2} + \mathbf{b} \left(\delta \hat A\right)^{2} + \mathbf{c}\left(\frac{\delta \hat \phi}{\hat \phi} \delta \hat A\right)\right) \ea
Where we have defined
\ba
\label{parameter}
Y&&\equiv \left( \frac{g \cos{\gamma}}{3\kappa \mathbf{e}}e^{-3N} - \frac{p_{c}\tan{\gamma}}{2}\right) \nonumber\\
\mathbf{a}&& \equiv -\frac{p_{c}}{4Y^{3}}\left((\frac{1}{3}+\frac{p_{c}}{2})\frac{g^2 \cos^2 \gamma}{3\mathbf{e}^2 \kappa^2}e^{-6N}- \frac{p_{c}g \sin \gamma}{2\mathbf{e}\kappa}e^{-3N}- (\frac{p_{c}}{2})^{2} \right)\nonumber\\
\mathbf{b}&& \equiv \frac{e^{-6N}}{6\kappa^2 Y^{3}}\left((\frac{1}{3}-\frac{p_{c}}{2})(\frac{p_{c}\tan^2 \gamma}{2})+ \frac{g \sin \gamma}{9\mathbf{e}\kappa}e^{-3N} + \frac{p_{c}}{6}\right)\nonumber\\
\mathbf{c}&& \equiv \frac{p_{c}e^{-3N}}{3\kappa Y^{3}}\left((\frac{1}{3}-\frac{p_{c}}{2})(\frac{g\sin \gamma}{2\mathbf{e}\kappa})e^{-3N}+ \frac{p_{c}}{4 \cos^{2}\gamma}\right)\nonumber\\
\ea
Now by substituting  Eqs. (\ref{delta1}) and  (\ref{delta2}) into Eq. (\ref{delta N phi}), it is straightforward to calculate $\delta N$
\ba
\label{deltaNN}
-\frac{2}{p_{c}}\delta N(\hat \phi,\hat A) = {\cal A}~\frac{\delta \hat \phi}{\hat \phi} + {\cal N}~\delta \hat A + {\cal I}~ \left(\frac{\delta \hat \phi}{\hat \phi}\right)^{2} +  {\cal S}~\left(\delta \hat A\right)^{2} + {\cal T}~\left( \frac{\delta \hat \phi}{\hat \phi}\delta \hat A \right)
\ea
where we have defined
\ba
{\cal A}&&\equiv \left( \frac{{g \cos{\gamma}}}{3\kappa \mathbf{e}} \frac{ e^{-3N}}{Y}\right) \nonumber\\
{\cal N}&&\equiv \left( \frac{{\tan{\gamma}}}{3\kappa} \frac{ e^{-3N}}{Y}\right) \nonumber\\
{\cal I}&&\equiv \left( \frac{p_{c}^{2} g}{24 \mathbf{e}\kappa} (\frac{1 + 4 \sin^2{\gamma}}{\cos{\gamma}}) \frac{ e^{-3N}}{Y^{3}}+ \frac{p_{c} g^2 }{12 \mathbf{e}^{2}\kappa ^{2}}\sin {2\gamma} (\frac{1}{3} - \frac{p_{c}}{4})\frac{ e^{-6N}}{Y^{3}} - \frac{g^3}{54 \mathbf{e}^{3}\kappa ^{3}} \cos^{3}{\gamma} \frac{ e^{-9N}}{Y^{3}}\right) \nonumber\\
{\cal S}&&\equiv \left( -\frac{p_{c}^{2}}{24 \kappa^2}\tan^3\gamma \frac{ e^{-6N}}{Y^{3}}+ \frac{g}{54 \mathbf{e} \kappa ^{3}} (\frac{1 +  \sin^2{\gamma}}{\cos{\gamma}}) \frac{ e^{-9N}}{Y^{3}}\right) \nonumber\\
{\cal T}&&\equiv \left( \frac{p_{c} g}{6\mathbf{e} \kappa^2}\left[\frac{((\frac{1}{3} - \frac{p_{c}}{2})\sin^2{\gamma} + \frac{1}{3})}{\cos{\gamma}}\right] \frac{e^{-6N}}{Y^{3}}\right) \, .\nonumber\\
\ea



\section*{References}
\begin{thebibliography}{}


\bibitem{Eriksen:2003db}
  H.~K.~Eriksen, F.~K.~Hansen, A.~J.~Banday, K.~M.~Gorski and P.~B.~Lilje,
  ``Asymmetries in the CMB anisotropy field,''
  Astrophys.\ J.\  {\bf 605}, 14 (2004)
  [Erratum-ibid.\  {\bf 609}, 1198 (2004)]
  [arXiv:astro-ph/0307507].
  
  F.~K.~Hansen, A.~J.~Banday and K.~M.~Gorski,
  ``Testing the cosmological principle of isotropy: local power spectrum
  estimates of the WMAP data,''
  Mon.\ Not.\ Roy.\ Astron.\ Soc.\  {\bf 354}, 641 (2004)
  [arXiv:astro-ph/0404206].
  
  T.~R.~Jaffe, A.~J.~Banday, H.~K.~Eriksen, K.~M.~Gorski and F.~K.~Hansen,
  ``Evidence of vorticity and shear at large angular scales in the WMAP  data:
  A violation of cosmological isotropy?,''
  Astrophys.\ J.\  {\bf 629}, L1 (2005)
  [arXiv:astro-ph/0503213].
  
  H.~K.~Eriksen, A.~J.~Banday, K.~M.~Gorski, F.~K.~Hansen and P.~B.~Lilje,
  ``Hemispherical power asymmetry in the three-year Wilkinson Microwave
  Anisotropy Probe sky maps,''
  Astrophys.\ J.\  {\bf 660}, L81 (2007)
  [arXiv:astro-ph/0701089];
  
  A.~de Oliveira-Costa, M.~Tegmark, M.~Zaldarriaga and A.~Hamilton,
  ``The significance of the largest scale CMB fluctuations in WMAP,''
  Phys.\ Rev.\  D {\bf 69}, 063516 (2004)
  [arXiv:astro-ph/0307282].
  
  K.~Land and J.~Magueijo,
  ``The axis of evil,''
  Phys.\ Rev.\ Lett.\  {\bf 95}, 071301 (2005)
  [arXiv:astro-ph/0502237].
  
  K.~Land and J.~Magueijo,
  ``The Axis of Evil revisited,''
  Mon.\ Not.\ Roy.\ Astron.\ Soc.\  {\bf 378}, 153 (2007)
  [arXiv:astro-ph/0611518].
  
  C.~Copi, D.~Huterer, D.~Schwarz and G.~Starkman,
  ``The Uncorrelated Universe: Statistical Anisotropy and the Vanishing Angular
  Correlation Function in WMAP Years 1-3,''
  Phys.\ Rev.\  D {\bf 75}, 023507 (2007)
  [arXiv:astro-ph/0605135].
  
  J.~Hoftuft, H.~K.~Eriksen, A.~J.~Banday, K.~M.~Gorski, F.~K.~Hansen and P.~B.~Lilje,
  ``Increasing evidence for hemispherical power asymmetry in the five-year WMAP
  data,''
  Astrophys.\ J.\  {\bf 699}, 985 (2009)
  [arXiv:0903.1229 [astro-ph.CO]].
  
  P.~K.~Samal, R.~Saha, P.~Jain and J.~P.~Ralston,
  ``Signals of Statistical Anisotropy in WMAP Foreground-Cleaned Maps,''
  Mon.\ Not.\ Roy.\ Astron.\ Soc.\  {\bf 396}, 511 (2009)
  [arXiv:0811.1639 [astro-ph]].
  
  A.~Hajian and T.~Souradeep,
  ``Measuring Statistical isotropy of the CMB anisotropy,''
  Astrophys.\ J.\  {\bf 597}, L5 (2003)
  [arXiv:astro-ph/0308001].
  
  A.~Hajian and T.~Souradeep,
  ``Testing Global Isotropy of Three-Year Wilkinson Microwave Anisotropy Probe
  (WMAP) Data: Temperature Analysis,''
  Phys.\ Rev.\  D {\bf 74}, 123521 (2006)
  [arXiv:astro-ph/0607153].

\bibitem{Komatsu:2010fb}
  E.~Komatsu {\it et al.},
  ``Seven-Year Wilkinson Microwave Anisotropy Probe (WMAP) Observations:
  Cosmological Interpretation,''
  arXiv:1001.4538 [astro-ph.CO].


\bibitem{Hanson:2009gu}
  D.~Hanson, A.~Lewis,
  ``Estimators for CMB Statistical Anisotropy,''
  Phys.\ Rev.\  {\bf D80}, 063004 (2009).
  [arXiv:0908.0963 [astro-ph.CO]].

\bibitem{Hanson:2010gu}
  D.~Hanson, A.~Lewis, A.~Challinor,
  ``Asymmetric Beams and CMB Statistical Anisotropy,''
  Phys.\ Rev.\  {\bf D81}, 103003 (2010).
  [arXiv:1003.0198 [astro-ph.CO]].


\bibitem{Ford:1989me}
  L.~H.~Ford,
  ``INFLATION DRIVEN BY A VECTOR FIELD,''
  Phys.\ Rev.\  D {\bf 40}, 967 (1989).

\bibitem{Kaloper:1991rw}
  N.~Kaloper,
  ``Lorentz Chern-Simons terms in Bianchi cosmologies and the cosmic no hair
  conjecture,''
  Phys.\ Rev.\  D {\bf 44}, 2380 (1991).
  
\bibitem{Kawai:1998bn}
  S.~Kawai and J.~Soda,
  ``Non-singular Bianchi type I cosmological solutions from 1-loop  superstring
  effective action,''
  Phys.\ Rev.\  D {\bf 59}, 063506 (1999)
  [arXiv:gr-qc/9807060].
  
\bibitem{Barrow:2005qv}
  J.~D.~Barrow and S.~Hervik,
  ``Anisotropically inflating universes,''
  Phys.\ Rev.\  D {\bf 73}, 023007 (2006)
  [arXiv:gr-qc/0511127].

\bibitem{Barrow:2009gx}
  J.~D.~Barrow and S.~Hervik,
  ``Simple Types of Anisotropic Inflation,''
  Phys.\ Rev.\  D {\bf 81}, 023513 (2010)
  [arXiv:0911.3805 [gr-qc]].
  
\bibitem{Campanelli:2009tk}
  L.~Campanelli,
  ``A Model of Universe Anisotropization,''
  Phys.\ Rev.\  D {\bf 80}, 063006 (2009)
  [arXiv:0907.3703 [astro-ph.CO]].
  
\bibitem{Golovnev:2008cf}
  A.~Golovnev, V.~Mukhanov and V.~Vanchurin,
  ``Vector Inflation,''
  JCAP {\bf 0806}, 009 (2008)
  [arXiv:0802.2068 [astro-ph]];
  
\bibitem{Kanno:2008gn}
  S.~Kanno, M.~Kimura, J.~Soda and S.~Yokoyama,
  ``Anisotropic Inflation from Vector Impurity,''
  JCAP {\bf 0808}, 034 (2008).

\bibitem{Pitrou:2008gk}
  C.~Pitrou, T.~S.~Pereira, J.~-P.~Uzan,
  ``Predictions from an anisotropic inflationary era,''
  JCAP {\bf 0804}, 004 (2008).
  [arXiv:0801.3596 [astro-ph]];
  T.~S.~Pereira, C.~Pitrou, J.~-P.~Uzan,
  ``Theory of cosmological perturbations in an anisotropic universe,''
  JCAP {\bf 0709}, 006 (2007).
  [arXiv:0707.0736 [astro-ph]].
  
\bibitem{Moniz:2010cm}
  P.~V.~Moniz, J.~Ward,
 ``Gauge field back-reaction in Born Infeld cosmologies,''
  [arXiv:1007.3299 [gr-qc]].

\bibitem{Boehmer:2007ut}
  C.~G.~Boehmer, D.~F.~Mota,
  ``CMB Anisotropies and Inflation from Non-Standard Spinors,''
  Phys.\ Lett.\  {\bf B663}, 168-171 (2008).
  [arXiv:0710.2003 [astro-ph]].


\bibitem{Koivisto:2008xf}
  T.~S.~Koivisto, D.~F.~Mota,
  ``Vector Field Models of Inflation and Dark Energy,''
  JCAP {\bf 0808}, 021 (2008).
  [arXiv:0805.4229 [astro-ph]].

\bibitem{Golovnev:2011yc}
  A.~Golovnev,
  ``On cosmic inflation in vector field theories,'' [arXiv:1109.4838 [gr-qc]].

\bibitem{Maleknejad:2011jr}
  A.~Maleknejad, M.~M.~Sheikh-Jabbari, J.~Soda,
  ``Gauge-flation and Cosmic No-Hair Conjecture,''
   [arXiv:1109.5573 [hep-th]].

  
\bibitem{Ackerman:2007nb}
  L.~Ackerman, S.~M.~Carroll and M.~B.~Wise,
  ``Imprints of a Primordial Preferred Direction on the Microwave Background,''
  Phys.\ Rev.\  D {\bf 75}, 083502 (2007).
  
  \bibitem{Yokoyama:2008xw}
  S.~Yokoyama and J.~Soda,
  ``Primordial statistical anisotropy generated at the end of inflation,''
  JCAP {\bf 0808}, 005 (2008);

\bibitem{Dimopoulos:2009vu}
 K.~Dimopoulos, D.~H.~Lyth and Y.~Rodriguez,
  ``Statistical anisotropy of the curvature perturbation from vector field
  perturbations,''
  arXiv:0809.1055 [astro-ph];
  
  T.~Kahniashvili, G.~Lavrelashvili and B.~Ratra,
  ``CMB Temperature Anisotropy from Broken Spatial Isotropy due to an
  Homogeneous Cosmological Magnetic Field,''
  Phys.\ Rev.\  D {\bf 78}, 063012 (2008);
  
  K.~Dimopoulos, M.~Karciauskas and J.~M.~Wagstaff,
  ``Vector Curvaton without Instabilities,''
  Phys.\ Lett.\  B {\bf 683}, 298 (2010)
  [arXiv:0909.0475 [hep-ph]];

  M.~Karciauskas, K.~Dimopoulos, D.~H.~Lyth,
  ``Anisotropic non-Gaussianity from vector field perturbations,''
  Phys.\ Rev.\  {\bf D80}, 023509 (2009).
  [arXiv:0812.0264 [astro-ph]];
  
  M.~Karciauskas,
  ``The Primordial Curvature Perturbation from Vector Fields of General non-Abelian Groups,''  [arXiv:1104.3629 [astro-ph.CO]].
  
  K.~Dimopoulos, M.~Karciauskas and J.~M.~Wagstaff,
  ``Vector Curvaton with varying Kinetic Function,''
  Phys.\ Rev.\ D {\bf 81}, 023522 (2010)
  [arXiv:0907.1838 [hep-ph]].


\bibitem{Dimastrogiovanni:2010sm}
  E.~Dimastrogiovanni, N.~Bartolo, S.~Matarrese and A.~Riotto,
  ``Non-Gaussianity and statistical anisotropy from vector field populated
  inflationary models,''
  arXiv:1001.4049 [astro-ph.CO].
    
\bibitem{ValenzuelaToledo:2009af}
  C.~A.~Valenzuela-Toledo, Y.~Rodriguez and D.~H.~Lyth,
  ``Non-gaussianity at tree- and one-loop levels from vector field
  perturbations,''
  Phys.\ Rev.\  D {\bf 80}, 103519 (2009)
  [arXiv:0909.4064 [astro-ph.CO]];
  
  C.~A.~Valenzuela-Toledo and Y.~Rodriguez,
  ``Non-gaussianity from the trispectrum and vector field perturbations,''
  Phys.\ Lett.\  B {\bf 685}, 120 (2010)
  [arXiv:0910.4208 [astro-ph.CO]].


\bibitem{Himmetoglu:2008zp}
  B.~Himmetoglu, C.~R.~Contaldi, M.~Peloso,
  ``Instability of anisotropic cosmological solutions supported by vector fields,''
  Phys.\ Rev.\ Lett.\  {\bf 102}, 111301 (2009).
  [arXiv:0809.2779 [astro-ph]];
  B.~Himmetoglu, C.~R.~Contaldi, M.~Peloso,
  ``Instability of the ACW model, and problems with massive vectors during inflation,''
  Phys.\ Rev.\  {\bf D79}, 063517 (2009).
  [arXiv:0812.1231 [astro-ph]];
  B.~Himmetoglu, C.~R.~Contaldi and M.~Peloso,
  ``Ghost instabilities of cosmological models with vector fields nonminimally
  coupled to the curvature,''
  Phys.\ Rev.\  D {\bf 80}, 123530 (2009)
  [arXiv:0909.3524 [astro-ph.CO]].







\bibitem{Martin:2007ue}
  J.~Martin and J.~Yokoyama,
  ``Generation of Large-Scale Magnetic Fields in Single-Field Inflation,''
  JCAP {\bf 0801}, 025 (2008)
  [arXiv:0711.4307 [astro-ph]].

\bibitem{Demozzi:2009fu}
  V.~Demozzi, V.~Mukhanov, H.~Rubinstein,
  ``Magnetic fields from inflation?,''
  JCAP {\bf 0908}, 025 (2009).
  [arXiv:0907.1030 [astro-ph.CO]].


\bibitem{Emami:2009vd}
  R.~Emami, H.~Firouzjahi and M.~S.~Movahed,
  ``Inflation from Charged Scalar and Primordial Magnetic Fields?,''
  Phys.\ Rev.\  D {\bf 81}, 083526 (2010)
  [arXiv:0908.4161 [hep-th]].


\bibitem{Kanno:2009ei}
  S.~Kanno, J.~Soda and M.~a.~Watanabe,
  ``Cosmological Magnetic Fields from Inflation and Backreaction,''
  JCAP {\bf 0912}, 009 (2009)
  [arXiv:0908.3509 [astro-ph.CO]].
  
\bibitem{Urban:2011bu}
  F.~R.~Urban,
  ``On inflating magnetic fields, and the backreactions thereof,'' [arXiv:1111.1006 [astro-ph.CO]].

\bibitem{Watanabe:2009ct}
  M.~a.~Watanabe, S.~Kanno and J.~Soda,
  ``Inflationary Universe with Anisotropic Hair,''
  Phys.\ Rev.\ Lett.\  {\bf 102}, 191302 (2009)
  [arXiv:0902.2833 [hep-th]].

\bibitem{Emami:2010rm}
  R.~Emami, H.~Firouzjahi, S.~M.~Sadegh Movahed, M.~Zarei,
  ``Anisotropic Inflation from Charged Scalar Fields,''
  JCAP {\bf 1102 } (2011)  005.
  [arXiv:1010.5495 [astro-ph.CO]].

\bibitem{Kanno:2010nr}
  S.~Kanno, J.~Soda, M.~-a.~Watanabe,
  ``Anisotropic Power-law Inflation,''
  JCAP {\bf 1012}, 024 (2010).
  [arXiv:1010.5307 [hep-th]].

\bibitem{Murata:2011wv}
  K.~Murata, J.~Soda,
  ``Anisotropic Inflation with Non-Abelian Gauge Kinetic Function,''
  JCAP {\bf 1106}, 037 (2011).
  [arXiv:1103.6164 [hep-th]].


\bibitem{Bhowmick:2011em}
  S.~Bhowmick, S.~Mukherji,
  ``Anisotropic Power Law Inflation from Rolling Tachyons,''
  [arXiv:1105.4455 [hep-th]].

\bibitem{Hervik:2011xm}
  S.~Hervik, D.~F.~Mota, M.~Thorsrud,
  ``Inflation with stable anisotropic hair: Is it cosmologically viable?,'' 
   [arXiv:1109.3456 [gr-qc]].


\bibitem{Dimopoulos:2010xq} 
  J.~M.~Wagstaff and K.~Dimopoulos,
  ``Particle Production of Vector Fields: Scale Invariance is Attractive,''
  Phys.\ Rev.\ D {\bf 83}, 023523 (2011)
  [arXiv:1011.2517 [hep-ph]].




\bibitem{Dulaney:2010sq}
  T.~R.~Dulaney, M.~I.~Gresham,
  ``Primordial Power Spectra from Anisotropic Inflation,''
  Phys.\ Rev.\  {\bf D81}, 103532 (2010).
  [arXiv:1001.2301 [astro-ph.CO]].


\bibitem{Gumrukcuoglu:2010yc}
  A.~E.~Gumrukcuoglu, B.~Himmetoglu, M.~Peloso,
  ``Scalar-Scalar, Scalar-Tensor, and Tensor-Tensor Correlators from Anisotropic Inflation,''
  Phys.\ Rev.\  {\bf D81}, 063528 (2010).
  [arXiv:1001.4088 [astro-ph.CO]].

  

\bibitem{Watanabe:2010fh}
  M.~a.~Watanabe, S.~Kanno and J.~Soda,
  ``The Nature of Primordial Fluctuations from Anisotropic Inflation,''
  Prog.\ Theor.\ Phys.\  {\bf 123}, 1041 (2010)
  [arXiv:1003.0056 [astro-ph.CO]].

\bibitem{Pereira:2007yy}
  T.~S.~Pereira, C.~Pitrou, J.~-P.~Uzan,
  ``Theory of cosmological perturbations in an anisotropic universe,''
  JCAP {\bf 0709}, 006 (2007).
  [arXiv:0707.0736 [astro-ph]];
  C.~Pitrou, T.~S.~Pereira, J.~-P.~Uzan,
  ``Predictions from an anisotropic inflationary era,''
  JCAP {\bf 0804}, 004 (2008).
  [arXiv:0801.3596 [astro-ph]].
  
\bibitem{Lyth:2005qk}
  D.~H.~Lyth,
  ``Generating the curvature perturbation at the end of inflation,''
  JCAP {\bf 0511}, 006 (2005).
  [astro-ph/0510443].



\bibitem{Linde:1993cn}
  A.~D.~Linde,
  ``Hybrid inflation,''
  Phys.\ Rev.\  D {\bf 49}, 748 (1994)
  [arXiv:astro-ph/9307002].
  
\bibitem{Copeland:1994vg}
  E.~J.~Copeland {\it et al.},
  ``False vacuum inflation with Einstein gravity,''
  Phys.\ Rev.\  D {\bf 49}, 6410 (1994)
  [arXiv:astro-ph/9401011].

\bibitem{Abolhasani:2010kr}
  A.~A.~Abolhasani, H.~Firouzjahi,
  ``No Large Scale Curvature Perturbations during Waterfall of Hybrid Inflation,''
  Phys.\ Rev.\  {\bf D83}, 063513 (2011).
  [arXiv:1005.2934 [hep-th]];

  A.~A.~Abolhasani, H.~Firouzjahi, M.~Sasaki,
  ``Curvature perturbation and waterfall dynamics in hybrid inflation,''
  JCAP {\bf 1110}, 015 (2011).
  [arXiv:1106.6315 [astro-ph.CO]].

\bibitem{Sasaki:1995aw}
  M.~Sasaki, E.~D.~Stewart,
  ``A General analytic formula for the spectral index of
 the density perturbations produced during inflation,''
  Prog.\ Theor.\ Phys.\  {\bf 95}, 71-78 (1996).
  [astro-ph/9507001].

 \bibitem{Wands:2000dp}
  D.~Wands, K.~A.~Malik, D.~H.~Lyth {\it et al.},
  ``A New approach to the evolution of cosmological perturbations on large scales,''
  Phys.\ Rev.\  {\bf D62}, 043527 (2000).
  [astro-ph/0003278].

\bibitem{Lyth:2005fi}
  D.~H.~Lyth, Y.~Rodriguez,
  ``The Inflationary prediction for primordial non-Gaussianity,''
  Phys.\ Rev.\ Lett.\  {\bf 95}, 121302 (2005).
  [astro-ph/0504045].

\bibitem{Sasaki:2008uc}
  M.~Sasaki,
  ``Multi-brid inflation and non-Gaussianity,''
  Prog.\ Theor.\ Phys.\  {\bf 120}, 159 (2008)
  [arXiv:0805.0974 [astro-ph]].

\bibitem{Naruko:2008sq}
  A.~Naruko and M.~Sasaki,
  ``Large non-Gaussianity from multi-brid inflation,''
  Prog.\ Theor.\ Phys.\  {\bf 121}, 193 (2009)
  [arXiv:0807.0180 [astro-ph]].


\bibitem{Huang:2009vk}
  Q.~-G.~Huang,
  ``A Geometric description of the non-Gaussianity generated at the end of multi-field inflation,''
  JCAP {\bf 0906}, 035 (2009).
  [arXiv:0904.2649 [hep-th]];

  Q.~-G.~Huang,
  ``The Trispectrum in the Multi-brid Inflation,''
  JCAP {\bf 0905}, 005 (2009).
  [arXiv:0903.1542 [hep-th]].



\bibitem{Hybrid}
  D.~H.~Lyth,
  ``Issues concerning the waterfall of hybrid inflation,''
  arXiv:1005.2461 [astro-ph.CO]; 

  J.~Fonseca, M.~Sasaki, D.~Wands,
``Large-scale Perturbations from the Waterfall Field in Hybrid Inflation,''
  JCAP {\bf 1009}, 012 (2010).
  [arXiv:1005.4053 [astro-ph.CO]]; 

  J.~-O.~Gong, M.~Sasaki,
  ``Waterfall field in hybrid inflation and curvature perturbation,''
  JCAP {\bf 1103}, 028 (2011).
  [arXiv:1010.3405 [astro-ph.CO]];

  D.~H.~Lyth,
  ``The contribution of the hybrid inflation waterfall to the 
 primordial curvature perturbation,''
   [arXiv:1012.4617 [astro-ph.CO]];


  L.~P.~Levasseur, G.~Laporte, R.~Brandenberger,
  ``Analytical Study of Mode Coupling in Hybrid Inflation,''
  Phys.\ Rev.\  {\bf D82}, 123524 (2010).
  [arXiv:1004.1425 [hep-th]];

  A.~A.~Abolhasani, H.~Firouzjahi, M.~H.~Namjoo,
  ``Curvature Perturbations and non-Gaussianities from Waterfall Phase Transition during Inflation,''
  Class.\ Quant.\ Grav.\  {\bf 28}, 075009 (2011).
  [arXiv:1010.6292 [astro-ph.CO]].















\end{thebibliography}
\end{document}